\documentclass[11pt,twoside]{article}
\usepackage{asp2004}
\usepackage{psfig}
\usepackage{epsf}
\usepackage{graphics}
\usepackage{lscape}
\begin{document}
\title{Revealing AGN by Polarimetry}
\author{Makoto Kishimoto$^1$, Robert Antonucci$^2$, Catherine Boisson$^3$, Omer Blaes$^2$}
\affil{$^1$Institute for Astronomy, University of Edinburgh, Blackford Hill,
Edinburgh EH9 3HJ, UK}
\affil{$^2$Department of Physics, University of California, Santa
Barbara, CA 93106, U.S.A.}
\affil{$^3$LUTH, FRE 2462 du CNRS, associ\'ee \`a l'Universit\'e Denis Diderot,
Observatoire de Paris, Section de Meudon, F--92195 Meudon Cedex, France}

\begin{abstract}

Polarimetric study in the UV and optical has been one of the keys to
reveal the structure and nature of active galactic nuclei (AGN).
Combined with the HST's high spatial resolution, it has directly
confirmed the predicted scattering geometry of $\sim$100 pc scale and
accurately located the hidden nuclear position in some nearby active
galaxies. Recently, we are using optical spectropolarimetry to reveal
the nature of the accretion flow in the central engine of quasars.  This
is to use polarized light to de-contaminate the spectrum and investigate
the Balmer edge spectral feature, which is otherwise buried under the
strong emission from the outer region surrounding the central engine.

\end{abstract}
\thispagestyle{plain}

\section{Introduction}

We will describe the UV/optical polarimetric contributions to the AGN
studies in two spatial scales, namely $\sim$100 pc and $\sim$0.001 pc
scale.  Broadly, active nuclei of galaxies consist of (1) the central
engine, which we refer here to the region emitting a strong UV/optical
continuum, probably of the scale of $\sim 100 R_S \sim 0.001$ pc
for the black hole mass $M_{\rm BH} = 10^8 M_{\odot}$ where
$R_S=2GM_{\rm BH}/c^2$;  (2) the broad-line region (BLR) of $\sim 0.1$
pc scale, emitting the broad permitted emission lines (FWHM $\sim$
5000km/s); (3) the narrow-line region (NLR) of $\ga 100$ pc scale,
emitting the narrow permitted and forbidden lines ($\sim$ 500km/s).
Generally, one observes both the BLR and NLR in type 1 AGNs, while only
NLR in type 2s.

\section{Inner Narrow-Line Region: $\sim$100 pc scale structure}

One of the most important results in understanding the structure of AGNs
is that at least some of the type 2 AGNs are indistinguishable from type
1s if they are looked in scattered light, or polarized flux spectra
\citep{An93}.  Based on this and many other facts, it is widely believed
that at least part of the type 1 and 2 distinction is simply
orientation: the central engine and BLR are surrounded by torus-like
obscuring matter ($\sim$1 pc scale), and our line of sight to type 2s
is edge-on so that we don't see the nucleus directly. However, light
escapes along the torus axis direction (and this is directly seen in
type 1s) and partly reflected above or below the torus.

The spectropolarimetry would suggest that the spatial scale of the
scattering region is rather comparable to that of the NLR: the continuum
and broad lines are polarized in the same way but narrow lines are often
polarized differently (and at a much lower level), meaning that the
scattering region directly seen in type 2s is well outside the scale of
the BLR but {\it not} the NLR\footnote[1]{The NLR low polarization can
also be affected by the transmission through aligned dust grains in the
host galaxy \citep[e.g.][]{Go92}. At least for NGC1068, however, the
scattering origin seems to be favored by the HST optical imaging
polarimetry data \citep{Ca95a}.}.

The HST UV imaging polarimetry has spatially resolved this scattering
region.  One clearest example is for the Seyfert 2 galaxy NGC1068 by
Capetti et al. (1995a,b).  Similar work has been done for other type 2s
\citep[e.g.][]{Ca96,Ki02}.  Highly polarized UV light with
centrosymmetric PA pattern is observed to be extended roughly in a cone
shape. This is seen in the inner NLR, so the size is consistent with the
spectropolarimetric suggestion above.  The hidden nuclear location can
be robustly determined as the symmetric center of this pattern, and no
bright UV/optical source is found at this position.  For NGC1068, there
was some deviation from the centrosymmetric pattern in the original map
of \citet{Ca95b}, but this was shown to be essentially entirely
instrumental and the hidden nuclear location was refined accordingly
\citep{Ki99}. This scattering center should be quite robust, and turned
out to be consistent with many subsequent high-resolution investigations
in the IR \citep[e.g.][]{Ga03,Bo00}.

In NGC 1068, there is evidence that the UV continuum is rather purely
dominated by scattered light \citep{An94,Tr95}, and nuclear scatterers
are thought to be electrons. In this case, since the morphology of the
scattering region is resolved to be quite clumpy, we can actually
estimate the viewing angle of each resolved clump from its observed UV
polarization, and construct a 3D structure map of the 100 pc scale
region \citep{Ki99}.

\begin{figure}[ht]
    \plottwo{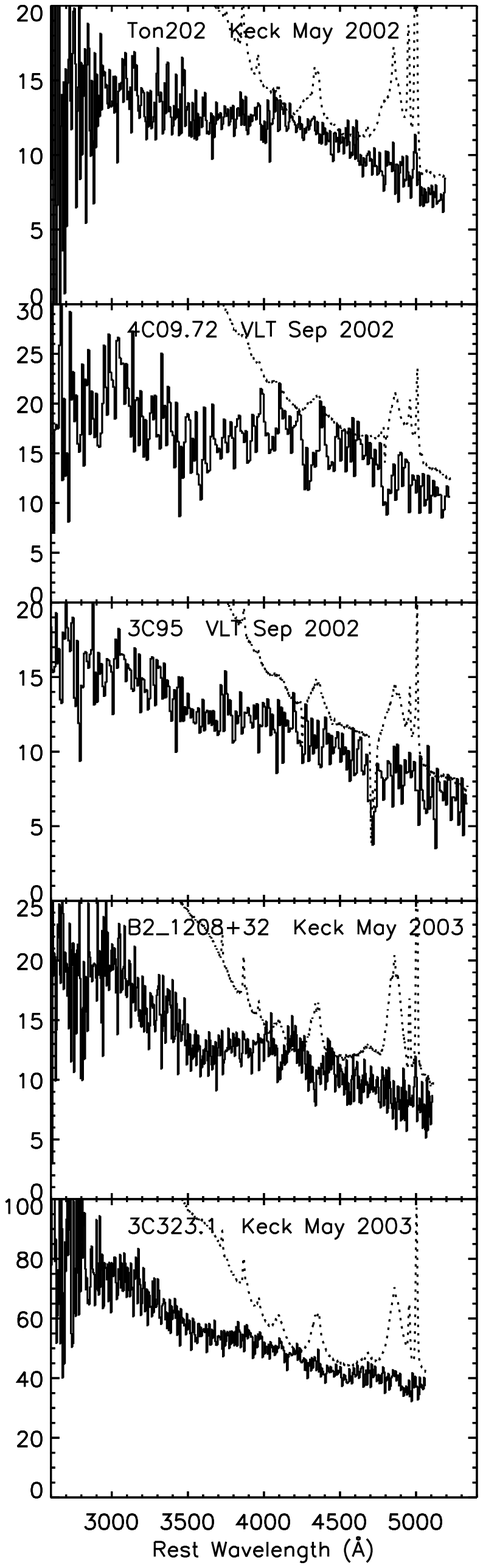}{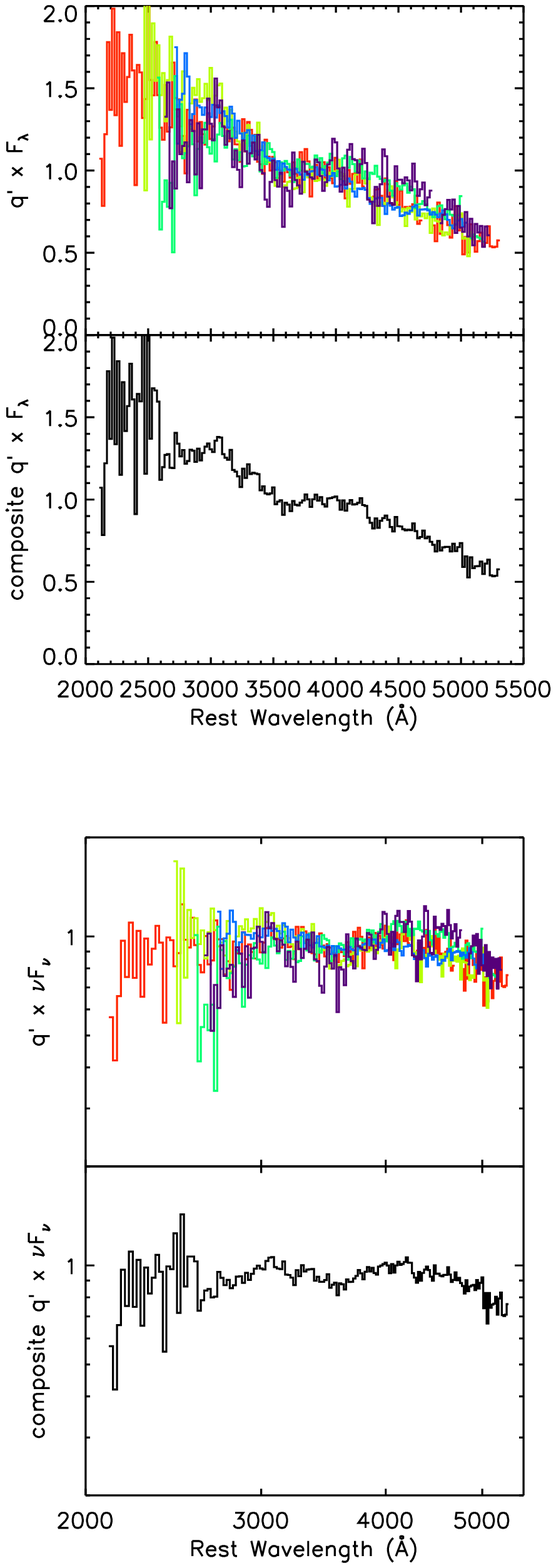}

    \caption{Our spectropolarimetric data for five quasars taken with
    the Keck and VLT \citep[from][]{Ki04}. Flux has been corrected
    for Galactic reddening. The wavelengths shown are all in
    the rest frame. Note that the data for all the objects except Ton
    202 and B2 1208+32 have been corrected for the interstellar
    polarization in our Galaxy. {\it Left}--- The solid line represents
    the polarized flux, and the dotted line is the total flux scaled to
    roughly match at the red side.  {\it Top-right}--- In the upper
    panel, the five polarized flux spectra, normalized by
    the mean at 2800-5000\AA, are over-plotted with a larger wavelength
    bin (20\AA). In the lower panel the average of these five spectra
    with equal weighting is shown. Both panels are in
    $F_{\lambda}$. {\it Bottom-right}--- The same as {\it top-right} but
    in $\nu F_{\nu}$ with both axes in log scale. } \label{fig-5obj}

\end{figure}

\section{Unresolved nucleus: $\sim$0.001 pc scale}

\subsection{The Balmer edge observed}

The central engine can't be spatially resolved yet, but polarimetry may
provide a very important clue to its understanding.  It has been well
known that many normal quasars (luminous type 1s) have 1\% level
polarization \citep[e.g.][]{St84,Be90}.  At least in some cases with
spectropolarimetry data of a limited S/N, this polarization has been
shown to be confined to continuum and not shared by emission from the
BLR (Antonucci 1988, showing the data of Miller and Goodrich; Schmidt \&
Smith 2000).  Our new data with a higher S/N now give a much tighter
limit on unpolarized broad lines.  Fig.\ref{fig-5obj} shows our 5 best
objects taken with the Keck and VLT.  Each panel shows the polarized
flux, with the dotted line being the total flux scaled to match the
polarized flux at the red side.  In the polarized flux, there is almost
nothing in the broad line wavelengths.

However, more importantly, the polarized flux shows definitely a
spectral feature in the Balmer edge wavelengths: in the wavelengths
longer than $\sim$4000\AA, the shape of the polarized flux is the same
as that of the total flux continuum, but it turns over at
$\sim$4000\AA, and then turns up below $\sim$3600\AA\ (thus, the
polarized flux has a local maximum at $\sim$4000\AA\ and a local minimum
at $\sim$3600\AA). This seems to be a Balmer edge feature seen in
absorption. In fact, this is actually what our observations were aiming
for, as we explain below.

\subsection{De-contaminating the spectra}

It is well known that the continuum emission from the central engine does
not seem to show a Lyman edge in general \citep*[e.g.][]{An89,Ko92}. The
UV/optical continuum is usually thought to be from an accretion disk
around the central supermassive black hole, and naively we should see a
Lyman edge feature, as many disk atmosphere models predict.  In some
individual cases and composites \citep[e.g.][]{Kr99,Zh97}, there seems
to be a slope change near the Lyman edge wavelength, but the theoretical
predictions are still not in a good agreement with these observations
\citep{Bl01}.  This apparent lack of the expected Lyman edge has been
one of the major difficulties in understanding the UV/optical continuum:
this component is energetically the most dominant, thus crucial to
understand.

One reason for the apparent lack could be that the Lyman edge is
relativistically smeared, since it is thought to originate from the
inner region of the disk. Lyman foreground absorption also
complicates/confuses the spectral region. On the other hand, the Balmer
edge is better in these two respects, since it would originate farther
out in radius and it is not a resonant feature; the only problem is
that the emission from the BLR heavily contaminates this spectral
region.

Our idea is to use the polarized flux to remove this contamination.  It
is simply based on the empirical fact that the polarization is confined
to the continuum and the emission from the BLR isn't polarized. Our
interpretation here is that the polarization is probably due to electron
scattering {\it interior to} the BLR\footnote[2]{We discuss and disfavor
the synchrotron possibility in \citet{Ki03}.  Scattering is supposed to
be by electrons and not by dust since it is within the dust sublimation
radius.}.  If the scattering region scale is much larger than the BLR,
then we would clearly see the broad lines polarized in the same way as
the continuum, just like the broad lines in type 2s.

Therefore, in the objects shown in Fig.\ref{fig-5obj} and possibly in
many other quasars, the polarized flux is likely to originate interior
to the BLR.  Then, by studying the polarized flux, we would be able to
effectively scrape off the contamination from the BLR and investigate
the BLR-emission-free continuum shape.  If this simple interpretation is
correct, the observed Balmer edge feature is attributed to be intrinsic
to the continuum emitter (the feature might be simply copied by
scattering, or polarization itself might be intrinsic to the emitter,
e.g. disk atmosphere), indicating that the emission is indeed thermal
and the emitter is optically thick. In addition, this opens up a totally
new way to look at the physical state of the accretion flow.

\subsection{Puzzles left}

However, we still cannot rule out the case where the Balmer edge
absorption signature is imprinted by the scatterers themselves
(``scatterer/absorber'' case). In fact, this is thought to be the case
in the circumstellar disks of Be stars (J. Bjorkman, K. Bjorkman in
these proceedings).  However, the fact that essentially no feature
(either in emission or absorption) is seen in the broad line wavelengths
seems to be against any simple scatterer/absorber models.  In addition,
the observed edge feature looks quite broadened, and this might require
quite specific conditions including a high velocity dispersion. Whatever
is producing the feature might thus be considered as a part of the
continuum emitter.  A detailed disk atmosphere model incorporating the
Balmer lines and metal lines is now being developed.

Many Seyfert 1 galaxies show broad lines which are polarized differently
from the continuum (J. Smith in these proceedings).  In this case, the
spatial scale of the scattering region is thought to be comparable to
that of the BLR.  These Seyfert galaxies are quite different in optical
luminosity and radio loudness from the objects in Fig.\ref{fig-5obj}
which are all radio-loud quasars. We note, however, that luminous
radio-loud type 1 object such as 3C382 also show polarized broad lines
like Seyfert 1s. Therefore, the distinction is not yet clear.

Finally, we note that there seems to be more spectral features in the
shorter wavelength region of the polarized flux, though they are less
clear: a local peak at $\sim$3050\AA\ and/or absorption feature
shortward of $\sim$3050\AA, and at $\sim$2600\AA. Those are clearer in
the composite spectrum which is constructed from the average of the 5
objects with equal weighting (see Fig.\ref{fig-5obj} and its
caption). The interpretation isn't clear yet, but they might be related
to FeII absorption.

\section{Conclusion}

Polarimetry has played a key role in understanding AGN. This has been
intensively done for type 2 objects, while polarization of type 1
objects is still under-exploited and has a potential to yield a very new
insight.  The Balmer edge feature observed and presented here is
entirely a new finding, and we definitely need a much larger sample to
investigate this further.


\begin{thebibliography}

\bibitem[Antonucci(1988)]{An88} Antonucci, R.\ 1988, 
Supermassive Black Holes, 26 

\bibitem[Antonucci, Kinney, \& Ford(1989)]{An89} Antonucci, 
R.~R.~J., Kinney, A.~L., \& Ford, H.~C.\ 1989, \apj, 342, 64 

\bibitem[Antonucci(1993)]{An93} Antonucci, R.\ 1993, \araa, 
31, 473 

\bibitem[Antonucci, Hurt, \& Miller(1994)]{An94} Antonucci, 
R., Hurt, T., \& Miller, J.\ 1994, \apj, 430, 210 

\bibitem[Berriman et al.(1990)Berriman, Schmidt, West, \& Stockman]{Be90} 
Berriman, G., Schmidt, G.~D., West, S.~C., \& Stockman, H.~S.\ 1990, \apjs, 
74, 869 

\bibitem[Blaes et al.(2001)Blaes, Hubeny, Agol, \& Krolik]{Bl01} Blaes, 
O., Hubeny, I., Agol, E., \& Krolik, J.~H.\ 2001, \apj, 563, 560 

\bibitem[Bock et al.(2000)]{Bo00} Bock, J.~J., et al.\ 2000, 
\aj, 120, 2904 

\bibitem[Capetti et al.(1995a)]{Ca95a} Capetti, A., Axon, 
D.~J., Macchetto, F., Sparks, W.~B., \& Boksenberg, A.\ 1995, \apj, 446, 
155 

\bibitem[Capetti et al.(1995b)]{Ca95b} Capetti, A., Macchetto, 
F., Axon, D.~J., Sparks, W.~B., \& Boksenberg, A.\ 1995, \apjl, 452, L87 

\bibitem[Capetti et al.(1996)]{Ca96} Capetti, A., Axon, 
D.~J., Macchetto, F., Sparks, W.~B., \& Boksenberg, A.\ 1996, \apj, 466, 
169 

\bibitem[Galliano et al.(2003)Galliano, Alloin, Granato, \& 
Villar-Mart{\'{\i}}n]{Ga03} Galliano, E., Alloin, D., 
Granato, G.~L., \& Villar-Mart{\'{\i}}n, M.\ 2003, \aap, 412, 615 

\bibitem[Goodrich(1992)]{Go92} Goodrich, R.~W.\ 1992, \apj, 
399, 50 

\bibitem[Kishimoto(1999)]{Ki99} Kishimoto, M. 1999, ApJ, 518, 676

\bibitem[Kishimoto et al.(2002)]{Ki02} Kishimoto, M., Kay, 
L.~E., Antonucci, R., Hurt, T.~W., Cohen, R.~D., \& Krolik, J.~H.\ 2002, 
\apj, 565, 155 

\bibitem[Kishimoto et al.(2003)Kishimoto, Antonucci, \& Blaes]{Ki03}
Kishimoto, M., Antonucci, R., \& Blaes, O. 2003, MNRAS, 345, 253

\bibitem[Kishimoto et al.(2004)]{Ki04}
Kishimoto, M., Antonucci, R., Boisson, C., \& Blaes, O. 2004, MNRAS, in press

\bibitem[Kriss et al.(1999)Kriss, Davidsen, Zheng, \& Lee]{Kr99} Kriss, 
G.~A., Davidsen, A.~F., Zheng, W., \& Lee, G.\ 1999, \apj, 527, 683 

\bibitem[Koratkar et al.(1992)Koratkar, Kinney, \& Bohlin]{Ko92} Koratkar, 
A.~P., Kinney, A.~L., \& Bohlin, R.~C.\ 1992, \apj, 400, 435 

\bibitem[Schmidt \& Smith(2000)]{SS00} Schmidt, G.~D.~\& 
Smith, P.~S.\ 2000, \apj, 545, 117 


\bibitem[Stockman et al.(1984)Stockman, Moore, \& Angel]{St84} Stockman, 
H.~S., Moore, R.~L., \& Angel, J.~R.~P.\ 1984, \apj, 279, 485 

\bibitem[Tran(1995)]{Tr95} Tran, H.~D.\ 1995, \apj, 440, 578 

\bibitem[Zheng et al.(1997)]{Zh97} Zheng, W., Kriss, G.~A., 
Telfer, R.~C., Grimes, J.~P., \& Davidsen, A.~F.\ 1997, \apj, 475, 469 





\end{thebibliography}
\end{document}